\documentclass[aps, two column, showpacs]{revtex4}

\begin{document}
\title{Do phonons carry spin?}
\author{S. C. Tiwari \\
Department of Physics, Institute of Science,  Banaras Hindu University,  and Institute of Natural Philosophy, \\
Varanasi 221005, India }
\begin{abstract}
Recently the question of phonon angular momentum has been raised : phonon angular momentum in a magnetic crystal could play an important role in the angular momentum balance in the Einstein-de Haas effect. This proposition is quite natural and understandable, however the idea of phonon spin seems ambiguous in the literature. In 1988 McLellan presented phonon angular momentum states for circular and elliptical polarizations of phonons that represent orbital angular momentum and also pointed out analogy to isotropic quantum harmonic oscillator. In view of its fundamental importance we investigate this issue and conclude that phonon spin hypothesis is physically untenable, and the phonon angular momentum has to be interpreted as orbital. The photon-phonon analogy is also misleading. It is argued that Cosserat elastic medium having internal torque may have lattice excitations different from phonons having spin one or spin two that we term as Cosseratons. In addition a new source of angular momentum balance originationg from anomalous electron magnetic moment is suggested.
\end{abstract}

\maketitle
\section{Introduction}

Recent advances in technology and material science have revived great interest in the interplay between mechanics and electromagnetism that has a long history \cite{1, 2}. Richardson in 1908 considered Ampere's molecular currents in nonmagnetized and magnetized states of a long thin iron rod, and suggested mechanical rotation induced by magnetization. Experimentally this effect was observed by Einstein and de Haas in 1915-1916; it seems appropriate to term it Richardson-Einstein-de Haas effect (RED-effect). An important question is raised in a recent paper \cite{3} as to the role of phonon angular momentum in the angular momentum balance in the RED-effect. Authors state that phonon angular momentum has never been discussed in the literature, however this is incorrect as already in 1988 phonon angular momentum was discussed by McLellan \cite{4}. Fundamental question of rotational symmetry in lattice dynamics and the conservation of the angular momentum motivated this work. Physical argument to ascribe angular momentum, not spin for phonons runs as follows. Linearly polarized phonons give zero average angular momentum, however the degenerate phonon states combined in circular or elliptical polarized basis give nonzero angular momentum. A remarkable result, Equation (5.2) in Section 5 of \cite{4} shows that the total angular momentum for the rotation of the equilibrium lattice comprises of phonon angular momentum, rigid-body lattice rotation and two small correction terms. Analogy to isotropic quantum harmonic oscillator angular momentum is also made; the phonon angular momentum is orbital.

Returning to \cite{3} the authors argue that rotating phonons could possess angular momentum that may have emergent macroscopic consequence, for example, in addition to the standard rigid body lattice angular momentum. They develop a microscopic theory for phonon angular momentum in a magnetic crystal recognizing the significance of the Raman spin-phonon interaction discussed in \cite{5}. Though the term phonon spin is not used anywhere in \cite{3} Garanin and Chudnovsky \cite{6} attribute the phonon spin concept to Zhang and Niu while revisiting the theory of spin relaxation. However, well established principles in classical and quantum field theories \cite{7, 8} show that there is no spin angular momentum for scalar field, and the separation to orbital and spin parts of the total angular momentum of the electromagnetic field (or photon) is not gauge invariant. Therefore a critical examination of the phonon spin hypothesis assumes great significance.

In the context of the RED-effect Zhang and Niu \cite{3} introduce a new term in the total angular momentum ${\bf J}^{tot}$ decomposition postulating nonzero phonon angular momentum ${\bf J}^{ph}$
\begin{equation}
{\bf J}^{tot} ={\bf J}^{lat} + {\bf J}^{ph} + {\bf J}^{orb} + {\bf J}^{spin}
\end{equation}
Imposing the condition of the angular momentum conservation one has
\begin{equation}
\Delta {\bf J}^{lat} + \Delta {\bf J}^{ph} + \Delta {\bf J}^{orb} + \Delta {\bf J}^{spin}=0
\end{equation}
On the other hand, the total magnetization change is given by
\begin{equation}
\Delta {\bf M} =\Delta {\bf M}^{orb} + \Delta {\bf M}^{spin}
\end{equation}
We know that the change in the magnetic moment is related with the change in the angular momentum, therefore we have
\begin{equation}
\Delta {\bf M}^{orb}= \frac{e}{2m c} \Delta {\bf J}^{orb}
\end{equation}
\begin{equation}
\Delta {\bf M}^{spin}= \frac{e}{m c} \Delta {\bf J}^{spin} 
\end{equation}
Authors \cite{3} make reasonable assumptions and use the experimental data to estimate the phonon angular momentum per unit cell of about $0.02 \hbar$ in paramagnetic $Ce F_3$, and $1.6 \times 10^{-4} \hbar$ in $Tb_3 Ga_5 O_{12}$. This being a very small value, it would be interesting to explore the possibility of another physical mechanism contributing to the balance angular momentum.

In this paper, we address the fundamental question whether the phonon spin hypothesis makes physical sense, and conclude that phonon spin concept is a flawed one. A detailed critique on this issue is presented in the next section. We discuss the implications of this conclusion, and suggest a new mechanism that may contribute to the angular momentum balance in the RED-effect  in Section III. The essential idea is very simple. Let us recall the QED value of the electron magnetic moment neglecting higher order terms
\begin{equation}
\mu_e =\mu_B (1+\frac{\alpha}{2 \pi})
\end{equation}
\begin{equation}
\mu_B = \frac{e \hbar}{2 m c}
\end{equation}
Here the fine structure constant $\alpha =\frac{e^2}{\hbar c}$. The expression (6) can be re-written in the form resembling the relation (5)
\begin{equation}
\mu_e = \frac{e}{mc} (\frac{\hbar}{2} + \frac{f}{2})
\end{equation}
\begin{equation}
f = \frac{e^2}{2 \pi c}
\end{equation}
Bohr magneton (7) or the first term on the right hand side of (8) signify the well-known interpretation that electron has the intrinsic spin of $\hbar /2$. It is logical and physically natural to attribute the complete term $[\frac{\hbar}{2} + \frac{f}{2}]$ in the expression (8) as the spin of the electron. Though this idea originated in our electron model \cite{9}, in the light of the discussion on the RED-effect following \cite{3}, we suggest that fractional spin in the unit of the Planck constant $\frac{f}{2} \frac{\hbar}{\hbar}=\frac{\alpha}{4 \pi} \hbar$ could be an important new ingredient in the angular momentum decomposition (1).

\section{Phonon spin hypothesis} 

Phonons are quantized lattice vibrations: in a three dimensional crystal with two atoms per unit cell  there are, in general, three acoustic and three optical branches. One may picture the three branches to correspond to one longitudinal and two transverse waves. The ensemble of n lattice vibrations in a line of n atoms could be treated formally as n independent harmonic oscillators. The average excitation energy of a lattice vibrational mode at a temperature T is given by
\begin{equation}
E_{av} = \frac{\hbar \omega}{2} + \frac{\hbar \omega}{(e^{\frac{\hbar \omega}{k T}} -1)}
\end{equation}
It may be noted that in the continuum limit of the atomic mass points phonons correspond to scalar wave, and a polarization index to label the three branches could be attached to the atomic displacement field.

In the nonrelativistic second quantized theory, a discrete excitation in the Fock space defines a phonon. One constructs occupation number space for n particles obeying Bose-Einstein statistics, and introduces creation and annihilation operators in the configuration space. A representation in which the number operator is diagonal defines a Fock space.

This preliminary account on the concept of phonons makes it easier to understand a physically meaningful description of the angular momentum of phonons in general, not necessarily restricted to a magnetic crystal \cite{3}. Two analogies are suggestive of phonon angular momentum. In the historically famous Newton's rotating water-bucket experiment we understand that solving the Euler's equation shows that the water rotates like a rigid body, after some time of setting the bucket into rotation, and the free water surface becomes a paraboloid. One may replace water by an ideal gas in a closed box: the rotation of the box imparts temperature dependent orbital angular momentum to the gas and variation in the pressure inside. The gas does not rotate like a rigid body. Imagine a phonon gas filled rotating box, in principle, there would exist a nonzero angular momentum of phonons.

Another analogy could be made with a relativistic massless scalar field since phonons are spin zero bosons. The Lagrangian density and the action for massless scalar field are invariant under infinitesimal Lorentz transformations, and there exists a conserved Noether current corresponding to this invariance. The time component of this conserved quantity is the orbital angular momentum density of the scalar field \cite{7}. Thus on physical grounds the existence of nonvanishing orbital angular momentum of phonons is natural.

Zhang and Niu \cite{3} do not use the word phonon spin in their paper, and define the angular momentum of phonons in terms of a displacement vector multiplied by square root of mass ${\bf u}_{la}$ of the a-th atom in the l-th cell
\begin{equation}
{\bf J}^{ph} = \sum_{la}{\bf u}_{la} \times \dot{{\bf u}}_{la}
\end{equation}
In fact, Sheng et al \cite{5} consider angular momentum ${\bf \Omega}_m$ and its correlation with phonon momentum, and tentatively state: "We may regard ${\bf \Omega}_m$ as an internal 'spin' degree of freedom of phonons." Now the question is: does expression (11) for phonon angular momentum represent spin?

It is straightforward to see from the expression
\begin{equation}
{\bf J}^{atom} = \sum_{la} ({\bf R}_{la} + {\bf u}_{la}) ~ \times (\dot{{\bf R}}_{la} + \dot{{\bf u}}_{la})
\end{equation}
given in \cite{3} that ${\bf J}^{ph}$ has the origin in the conventional rotation of unstructured points around the equilibrium position, not spin. Here ${\bf R}_{la}$ is the equilibrium position of the a-th atom in the l-th unit cell multiplied by the square root of its mass. The term $({\bf R}_{la} \times \dot{{\bf R}}_{la})$ in (12) would represent the rotation of the crystal. Since there is no nontrivial topological quantization involved in the phonon angular momentum one expects a continuum of values of orbital angular momentum per phonon in agreement with the calculated values in \cite{3}.

Preceding discussion shows that the phonon angular momentum considered by Zhang and Niu is not spin of phonons. Attributing phonon spin hypothesis to them as is done in \cite{6} is not correct, but then we must understand the theory of phonon angular momentum developed by Garanin and Chudnovsky \cite{6}.
Authors base their arguments in analogy to photon angular momentum and its separation into spin and orbital parts. It is claimed that the spin of a phonon is $\hbar$ for circularly polarized phonons. One could always envisage unconventional ideas in physics, however assuming the phonon to be a lattice vibration quantized mode in conformity with the textbooks, it being a scalar particle having zero spin, the photon analogy is misleading. A short discussion on the concept of photon \cite{10} points out that according to Mandel and Wolf \cite{11}: 'The discrete excitations or quanta of the electromagnetic field, corresponding to the occupation numbers $\{ n\}$, are usually known as photons. Thus a state $|0, 0, 0,...1_{k,s},0,..>$ is described as a state with one photon of wave vector ${\bf k}$ and polarization $s$'. Formally phonon and photon descriptions in the Fock space, and the appearance of a polarization index for both allow superficial justification for the analogy. However the physical content to polarization index $s$ relating it with spin crucially depends on the vector nature of the electromagnetic wave, and the well defined angular momentum density ${\bf r} \times ({\bf E} \times {\bf B})$ for photons. In fact, helicity rather than spin is more appropriate for the massless photon. Polarization index has symbolic significance for scalar sound waves, and cannot be identified representing phonon spin. Thus the phonon-photon analogy to establish phonon spin hypothesis is misleading.

Further, it has to be realized that even for the case of photon the separation of angular momentum into orbital and spin components remains a controversial and unsettled issue \cite{12} at a conceptual level. The question of canonical versus kinetic angular momentum and the role of gauge invariance in QED and QCD continue to be debated in the recent literature \cite{13}.

Interestingly the scalar field description of light not only proved useful in several optical phenomena in the past, recent studies on the orbital angular momentum of light beams also underline the utility of this approach \cite{14}. In this case there could be an unambiguous analogy to ascribe orbital angular momentum to phonons.

An important idea dependent on the model of an elastic medium possessing internal torque \cite{6} deserves serious attention. In this context we have some remarks. It is not clear if the magnetic crystal considered in \cite{3} requires the property that there exists internal torque. Moreover, is it physically justified to term the quantized lattice vibrations for such a medium as phonons? More than a century ago, Cosserat brothers \cite{15}  proposed a medium in which the displacement vector ${\bf a}_{la} $ is inadequate; an additional field is needed to describe rotation of the assumed structured points inside the medium. This additional field, similar to torsion,  gives rise to the antisymmetric spin stress tensor. In Cosserat medium one could envisage spin, but then the lattice vibrations are not the conventional phonons. How does the mediun considered in \cite{6} differ from the Cosserat medium?

The inevitable conclusion inferred from all these physical arguments is that phonons do not carry spin. Orbital angular momentum of phonons may be ubiquitous, and the angular momentum discussed by Zhang and Niu \cite{3} should be interpreted as orbital that has already been discussed by McLellan \cite{4}.

\section{Discussion and conclusion}

Electro- and magneto- mechanical systems at nano scale, spin-phonon interaction in magnetic materials, and phonon Hall effect are some of the important areas of current activities where the role of phonon angular momomentum could be of significant physical consequences \cite{3, 6}. The main conclusion arrived at in the previous section, namely, that only orbital angular momentum of phonons has a sound physical basis, assumes importance in this connection. Though phonon spin is argued to be unphysical, it is of interest to ask if 'intrinsic spin' for the lattice vibrations in an elastic medium having internal torque could exist. Note that the role of Cosserat medium \cite{15} in the development of the Einstein-Cartan spacetime has been recognized in the literature \cite{16}. However, we are not aware if quantized excitations in such a medium have been studied; the kind of elastic medium analyzed in \cite{6} appears essentially to be Cosserat medium \cite{15}. The existence of spin carrying lattice excitations in such a medium cannot be ruled out; however these cannot be identified as phonons. Let us term them Cosseratons. What is their spin? Garanin and Chudnovsky theory \cite{6} would seem to be the first attempt in this direction. Unfortunately, the spin component ${\L}^{(2)}$ interpreted as phonon spin lacks clarity because the second order in phonon operators \cite{6} does not imply vector nature of the phonon field. Once analogy to photon is assumed spin one follows trivially. We argue that the key element in a Cosserat medium is the bivector field $\omega^{ij} =-\omega^{ji}$ \cite{16}. Two logically possible approaches could be envisaged as follows. 1) Postulate a vector field to construct the bivector field, and in analogy to the electromagnetic potential quantize the field for spin one Cosseraton. And, 2) treat the bivector field as fundamental; in that case Cosseraton would be a spin two boson. In fact, a medium possessing internal torque deserves more attention for its potential to address some basic issues; for example, an alternative interpretation of the electromagnetic field tensor representing angular momentum per unit area for the photon fluid is suggested in \cite{10}.

Returning to the original motivation for phonon angular momentum \cite{3} let us ask a different question: could there be any role of anomalous magnetic moment in this connection? The role of electron magnetic moment and spin is obtained from Eqs. (2) and (5) where the anomalous part is neglected. In most of the condensed matter systems one could justifiably neglect the anomalous part in $\mu_e$. However authors \cite{3} in a remarkable result show that phonon angular momentum of the order of a small fraction of $\hbar$ may have observable effect in some physical situations. In such cases the effect of anomalous magnetic moment would be of comparable magnitude, and cannot be neglected. Note that in Dirac electron theory the electron spin $\frac{\hbar}{2}$ and corresponding spin magnetic moment $\mu_B$ emerge naturally. A new term introduced by Pauli modifies the magnetic moment, however in 1941 it was thought to be unnecessary for electron by Pauli \cite{17}. In 1947 Nafe et al experiment \cite{18} showed a difference of about 0.26 percent in the measured and calculated values of the hyperfine separation lines in hydrogen and deuterium indicating anomalous magnetic moment for the electron \cite{19}. Since the electron in this experiment is in a bound atomic state it may be argued that $\mu_e$ may not be a physical attribute of free electron. However a landmark experiment in 1986 \cite{20} measured the magnetic moment of free electron that differs from $\mu_B$; QED calculated value to order $\alpha$ is given by Eq.(6).

Bohr asserted that the determination of spin and spin magnetic moment for a free electron were two different issues, and the magnetic moment of a free electron could not be measured \cite{21, 22}. The later part of Bohr's assertion has been proved wrong by experiments \cite{20}, but the first part needs further discussion. The experiments on magnetization by rotation (Barnett effect) and rotation by magnetization (RED effect) \cite{1, 2} embody the principle of magneto-mechanical equivalence. Eqs.(3) and (4) exemplify this principle. Could we extend this principle for a free electron? Assuming its applicability to a free electron the rearranged form (8) of the expression (6) implies that electron spin should be $(\frac{\hbar}{2} +\frac{f}{2})$. Fractional spin for quasi-particles is accepted in the literature, however to ascribe fractional spin $\frac{f}{2}$ to a free electron is a radical departure. Though the electron model proposed in \cite{9, 23} is based on this hypothesis the contribution of this paper is independent of any such model: we suggest the importance of anomalous magnetic moment and associated spin in the magneto-mechanical effects in nano materials highlighted in \cite{3, 6}. In particular the angular momentum balance (1) and (2) must include electron fractional spin in the light of the sensitivity of measurements to detect phonon angular momentum as small as $10^{-4} \hbar$.

In conclusion, it has been shown that phonon spin hypothesis is unphysical, and photon-phonon analogy is misleading. It is suggested that lattice excitations of physical nature different than phonons, termed Cosseratons in a medium possessing internal torque, could be spin one or spin two bosons. The role of fractional spin associated with the anomalous part of electron magnetic moment in magneto-mechanical effects is proposed that seems testable in nano matter.

{\bf APPENDIX}

There are two questions raised in connection with this paper. We address them below:

I. Question: "The assumption that phonons are scalar excitations is incorrect; any condensed matter textbook teaches that phonons correspond to a quantized vector displacement field and have three polarizations"

It has been explained in the text above that displacement vector for lattice, in general is a vector, and the quasi-particles identified as phonons could be conveniently described using a polarization index; however the polarization index for phonons is different than photon. Note that it does not lead to spin of phonon, in fact, polarization index-spin relation is not a fundamental principle. An interesting paper \cite{24} shows that beginning with the quantized molecular field a quantum field theory of crystals could be developed and the classical theory could be obtained by a boson transformation method. Phonon spin does not arise here either in quantized theory or classical theory. Incidentally orbital angular momentum of phonons in analogy to optics was proposed in \cite{25}. 

II. Question: "In condensed matter the renormalization of the electron magnetic moment is much greater and is described by the gyromagnetic tensor specific to the material."

Not exactly, for most materials anomalous magnetic moment is insignificant and is not considered. It is only recently that new matter like Dirac and Weyl semimetal have led to calculations on anomalous magnetic moment in them For example, in \cite{26} van der Wurff and Stoof calculate it for massless electrons comparing with Schwinger's result. Therefore the question is misleading.

\end{document}